\documentclass[a4paper,11pt,onecolumn]{IEEEtran}
\usepackage{xurl}
\usepackage{pos}
\usepackage{xcolor}
\usepackage{caption}
\usepackage{subcaption}
\usepackage{filecontents} 
\usepackage{wrapfig}
\usepackage{siunitx}
\usepackage[left]{lineno}

\newcommand{\gphysnet}[0]{$\gamma$-PhysNet DA}

\title{Analysis of the Cherenkov Telescope Array first Large-Sized Telescope real data using convolutional neural networks}

\ShortTitle{CTA LST-1 real data analysis using CNNs}

\manuallySeparateAuthors
\author[a]{Thomas Vuillaume}
\author[a,b]{, Mika\"el Jacquemont}
\author[a]{, Mathieu de Bony de Lavergne}
\author[a]{, David A. Sanchez}
\author[a]{, Vincent Poireau}
\author[a]{, Gilles Maurin}
\author[b]{, Alexandre Benoit}
\author[b]{, Patrick Lambert}
\author[a]{, Giovanni Lamanna}
\renewcommand{\printHeadAuthors}{T.~Vuillaume et al. on behalf of the CTA-LST Collaboration}

\affiliation[a]{Laboratoire d’Annecy de Physique des Particules, Univ. Grenoble Alpes, Univ. Savoie Mont Blanc,\\
CNRS, LAPP, 9 chemin de bellevue, 74940 Annecy, France}

\affiliation[b]{LISTIC, Universit\'e Savoie Mont-Blanc Polytech\\ Annecy-Chamb\'ery, 5 chemin de bellevue, 74940 Annecy, France}

\forColl{CTA-LST} 

\emailAdd{thomas.vuillaume@lapp.in2p3.fr}

\abstract{
The Cherenkov Telescope Array (CTA) is the future ground-based gamma-ray observatory and will be composed of two arrays of imaging atmospheric Cherenkov telescopes (IACTs) located in the Northern and Southern hemispheres respectively. 
The first CTA prototype telescope built on-site, the Large-Sized Telescope (LST-1), is under commissioning in La Palma and has already taken data on numerous known sources. 
IACTs detect the faint flash of Cherenkov light indirectly produced after a very energetic gamma-ray photon has interacted with the atmosphere and generated an atmospheric shower. Reconstruction of the characteristics of the primary photons is usually done using a parameterization up to the third order of the light distribution of the images.
In order to go beyond this classical method, new approaches are being developed using state-of-the-art methods based on convolutional neural networks (CNN) to reconstruct the properties of each event (incoming direction, energy and particle type) directly from the telescope images. While promising, these methods are notoriously difficult to apply to real data due to differences (such as different levels of night sky background) between Monte Carlo (MC) data used to train the network and real data.
The GammaLearn project, based on these CNN approaches, has already shown an increase in sensitivity on MC simulations for LST-1 as well as a lower energy threshold. This work applies the GammaLearn network to real data acquired by LST-1 and compares the results to the classical approach that uses random forests trained on extracted image parameters. The improvements on the background rejection, event direction, and energy reconstruction are discussed in this contribution.
}

\FullConference{37$^{\rm{th}}$ International Cosmic Ray Conference (ICRC 2021)\\
		July 12th -- 23rd, 2021\\
		Online -- Berlin, Germany}

\begin{document}


\maketitle
\bstctlcite{IEEEexample:BSTcontrol}

\section{Introduction}


The Cherenkov Telescope Array (CTA) is the future of ground-based gamma-ray astronomy. With two arrays built on two sites, La Palma (Canary Islands) and Paranal (Chile), it will be composed of tens of imaging atmospheric Cherenkov telescopes (IACT) of different sizes that will increase the sensitivity by a factor of 5 to 10 compared to the current generation of instruments in the range from \SI{20}{\giga\electronvolt} to \SI{300}{\tera\electronvolt}.
The Large-Sized Telescope 1 (LST-1) is the first CTA prototype telescope built on-site (at the Observatorio del Roque de los Muchachos, La Palma) and is currently under commissioning. Thanks to its size (23 m mirror diameter), it is specially designed to focus on the low energy range of CTA (from \SI{20}{\giga\electronvolt} to a few hundreds of \SI{}{\giga\electronvolt}). Data from several known gamma-ray sources have already been acquired.

Relativistic particles (either gamma-rays or cosmic rays) entering the atmosphere produce atmospheric showers, whose sub-products produce Cherenkov light collected by IACTs. A complex analysis, also called \textit{event reconstruction}, is necessary to determine the particle type (separating gamma rays from cosmic rays which represent the background), its energy and incoming direction from the images produced by IACTs. We present this analysis in further details in section \ref{sec:event_reco} and then explain how it can benefit from convolutional neural networks (CNNs), comparing the standard approach and our network \gphysnet. We present performances on simulations in section \ref{sec:mc_perf} and on LST-1 data in section \ref{sec:lst1_data}.

\section{Event reconstruction \label{sec:event_reco}}

\begin{wrapfigure}{r}{0.45\textwidth}
  \begin{center}
  \vspace{-0.8cm}
    \includegraphics[width=0.3\textwidth]{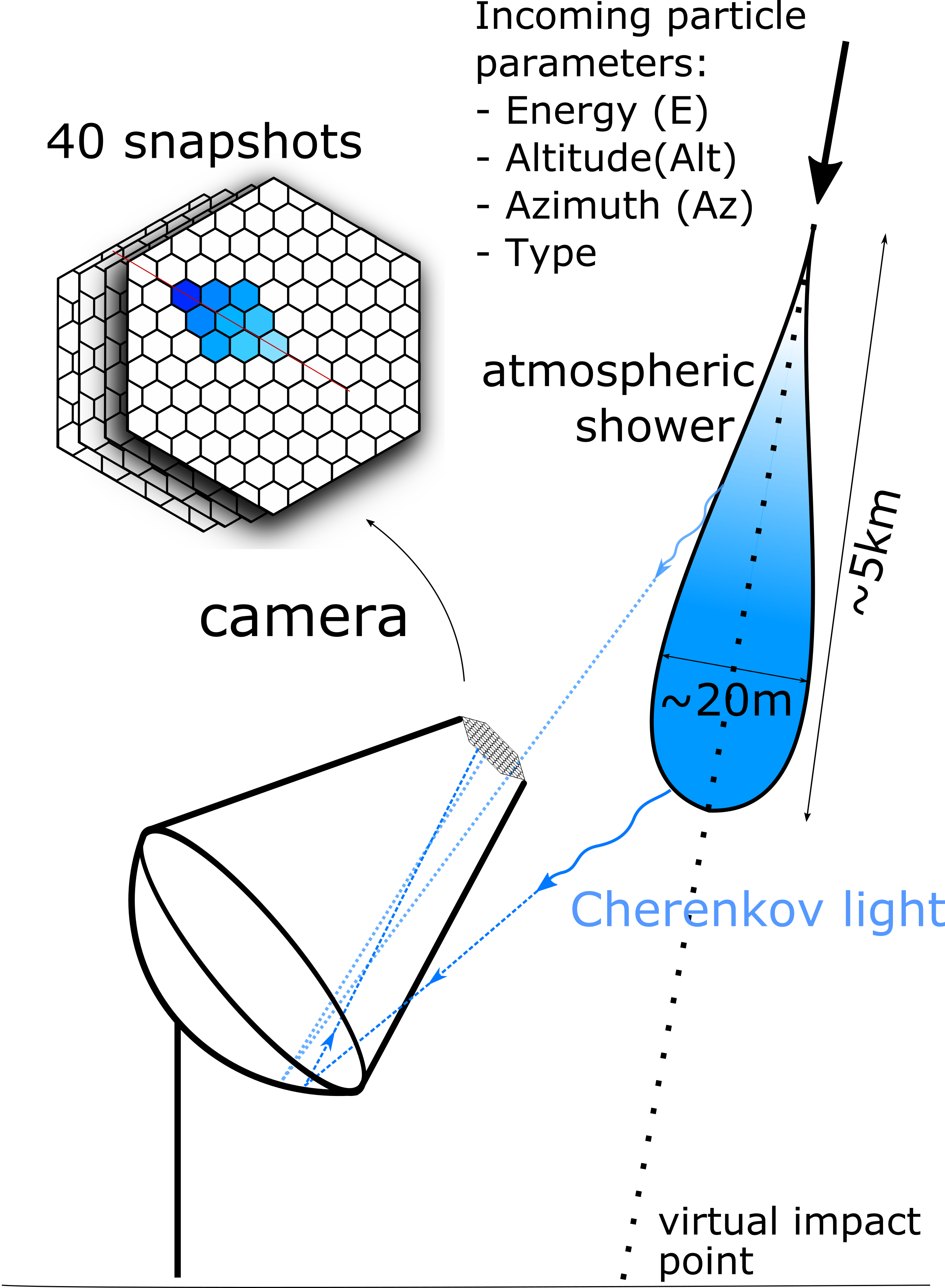}
    
    \vspace{0.25cm}
    
      \includegraphics[width=0.45\textwidth, trim={0 2cm 0 0cm}, clip]{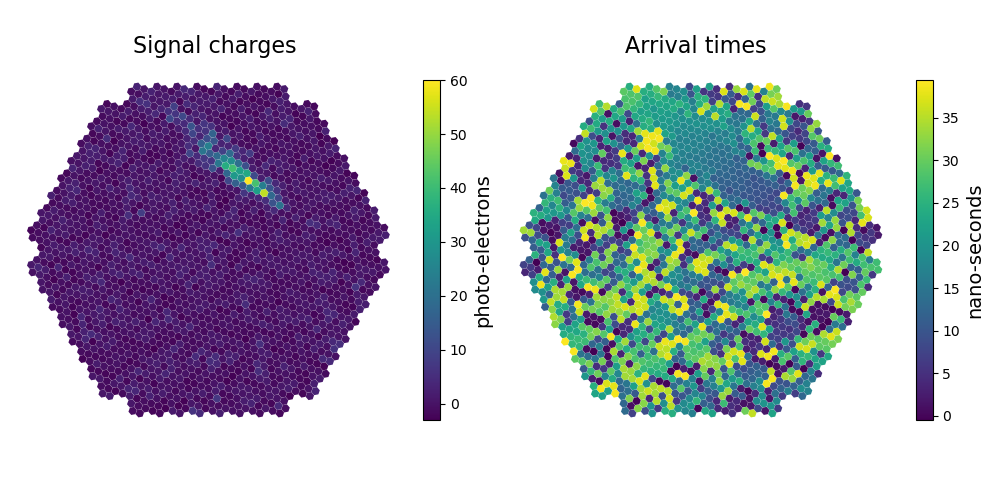}
     \end{center}
     
     \caption{Top: LST-1 observation principle. Bottom: Extracted signal charges and times.}
     \label{fig:iact}
  \vspace{-0.4cm}
\end{wrapfigure}

The Cherenkov light generated by atmospheric showers is collected on the ground by the LST-1 optical system that focuses it to an ultra-fast and sensitive camera sampling the signal into a temporal series of 40 snapshots of 1855 pixels of 1 ns each (also called waveform). This process is depicted in Figure \ref{fig:iact}.
The waveform is then calibrated and integrated to obtain two images, one containing the pixel charges and one the mean arrival time of the Cherenkov photon in each pixel. These two images are then analysed to extract image parameters used to reconstruct the particles physical parameters of interest (type, energy, incoming direction) using machine learning algorithms such as random forests \cite{random_forests}.


Thanks to accurate Monte Carlo simulations \cite{heck1998corsika, Bernlohr2008}, machine learning algorithms are trained on these images in order to reconstruct the corresponding physical parameters. The algorithms can then be applied to simulated test datasets in order to compute the instrument response function, and so the performance of the reconstruction, or to data from observations.

The steps of image parameters extraction, training, and inference on observations is the one we propose to enhance with deep learning. Indeed, since the AlexNet revolution \cite{10.1145/3065386} CNNs have shown in many fields \cite{2015arXiv151207108G} their capacity to replace and overtake solutions with engineered parameters extraction. They have been applied to IACT event reconstruction in the past on CTA simulated data \cite{nieto2017exploring, nieto2020adass, jacquemont2020adass} and on H.E.S.S. \cite{shilon2019application, 2020EPJC...80..363P} data, but with limited success on observations. 
Here we will compare the standard approach using engineered parameters and random forests with the CNN we developed specifically for the tasks at hand, \gphysnet.
The first step of calibration plus charge integration is common for both approaches and presented in further details in \cite{kobayashi_icrc2021}. It is performed using the lstchain v0.7.3 library \cite{lstchain_adass20} based on ctapipe v10.5 \cite{ctapipe105, pipelines-icrc2021}.

\subsection{LST-1 classical approach: Hillas and ensemble models}

A classical approach, called Hillas by the name of its creator \cite{Hillas1985}, is based on characteristic parameters extracted from these images. This is the standard analysis method in use for LST-1 data analysis.
The images first go through a cleaning step called tailcut-cleaning requiring pixels to be above a given picture threshold and a neighbor above a second threshold. The same mask is then applied on the temporal map. Image parameters are then extracted from these cleaned images. The parameters (defined in lstchain v0.7.3) used in this work are: 
image intensity, ellipsoid width, length, position and orientation in the camera, second order moments, signal time development gradient and charges containment.

Three different random forests (RF) are then trained, one for each task: particle classification, energy reconstruction and direction reconstruction. The performances of these RFs are presented in section \ref{sec:mc_perf}. This approach, later noted \textit{Hillas+RF}, will serve as the reference for the rest of this work. The complete set of parameters is given in lstchain v0.7.3 standard configuration.

\subsection{$\gamma$-PhysNet - a convolutional neural network to analyse LST-1 data}

The CNN we developed, \gphysnet, is presented in depth in Ref.~\cite{jacquemont2021visapp, jacquemontCBMI_arxiv}. It is a multi-task network, so contrary to the classical approach and to previous works using CNNs, a single algorithm is trained to perform the full event reconstruction (particle classification, energy reconstruction and direction reconstruction). It is composed of a first convolution block common to all tasks and based on ResNet-56 \cite{resnet_cvpr} enhanced with indexed convolutions \cite{visapp19indexed} and Dual Attention mechanism \cite{sun2020saunet}. 

\section{Performance comparison on simulated data\label{sec:mc_perf}}


The Monte Carlo simulations, produced using CORSIKA  \cite{heck1998corsika} and simtel\_array \cite{Bernlohr2008} are fine-tuned to the LST-1 at a pointing of $20\,$\textdegree\ zenith in the South direction. The training dataset is composed of isotropic gamma rays and protons while the test dataset is composed of point-source gammas at an offset of 0.4\textdegree~to the camera center, and isotropic protons and electrons.

Quality cuts are applied to select events: the integrated charge after image cleaning must be greater than 50 photoelectrons, the fraction of that integrated charge in the outermost ring of pixels in the camera must be below 20\% and the image must pass a tailcut cleaning with a picture threshold of 6 photoelectrons and a neighbor threshold of 3 photoelectrons. Note however that the cleaning itself is applied only in the case of the Hillas+RF procedure, while the entire image information is kept in the case of \gphysnet.


\subsection{Instrument response functions}

\begin{figure}[h]
    \centering
    \includegraphics[width=\textwidth]{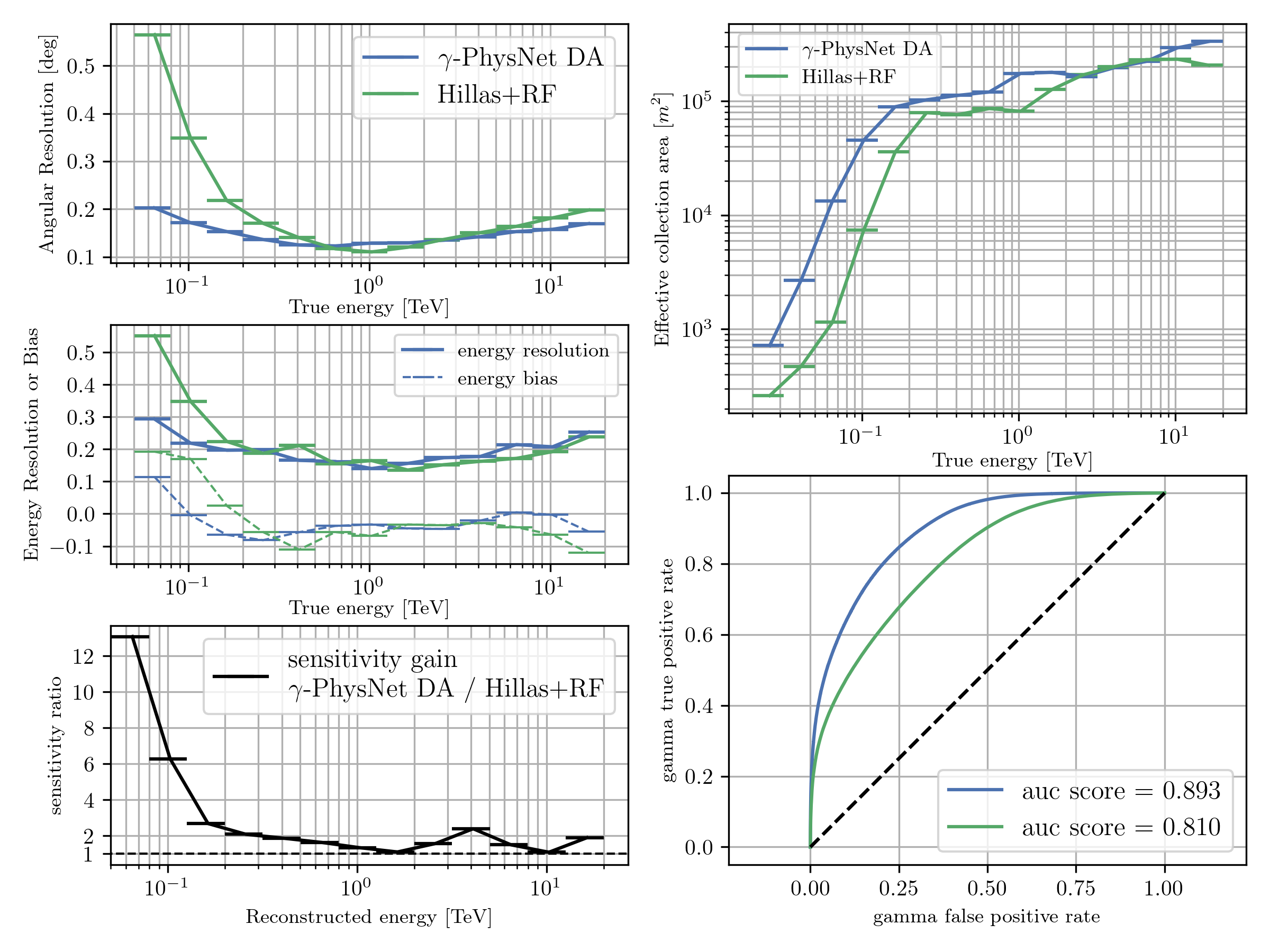}
    \caption{Comparison of instrument response functions and performances on simulated data between the reference (Hillas+RF) and our network \gphysnet.~Top-left panels: angular resolution, energy resolution and energy bias (lower is better). Bottom-left: Sensitivity ratio (higher is better). Top-right: effective area (higher is better). Bottom-right: ROC curve (higher auc score is better).}
    \label{fig:irfs}
\end{figure}

Applying the trained algorithms to the test dataset, one can compute the instrument response functions (Figure \ref{fig:irfs}, as defined in Ref. \cite{cta_perf_icrc2021}) for the two approaches. The gammaness and angular cuts are determined for best sensitivity in each energy bin using the pyirf v0.4 package \cite{pyirf_2020_4304466}. The highest performances of \gphysnet~on all tasks are clearly demonstrated at all energies and in particular at the lowest ones (below \SI{300}{\giga\electronvolt}). The greater classification power of \gphysnet~is shown by the higher effective area and by the higher area under the roc curve.

     

\section{Application to LST-1 data \label{sec:lst1_data}}



\subsection{Methodology\label{sec:methodo}}

The fairest way to compare the performances of two reconstruction chains is to match their background levels. To do so, we first apply the quality selection cuts common to both chains. Then, we fix the gammaness (likelihood of an event to be a gamma) cut of the standard Hillas+RF analysis at a given level and the gammaness cut of \gphysnet~is then determined in order to get the same background level as the one of the reference in each energy bin in the background region (determined with the multiple OFF technique \cite{multipleOFF_berge2007}). It is worth noticing that several values have been tested for the reference gammaness cut with no significant change in the results - only one analysis has been kept for simplicity.

\subsection{Detection of Markarian 501\label{sec:mrk501_results}}

Markarian 501 is a well-known blazar at \SI{}{\tera\electronvolt} energy range. The observation dataset is composed of 4 runs taken on the night of the 20 March 2021 for a total of 1h. Observed zenith angles range from 20° to 12.5°. All observations are conducted in wobble mode with an offset of 0.4°.

In this case, the analysis (training and inference) has been performed with cleaning levels of (8,4). The high-level analysis is performed using gammapy v0.18.2 \cite{gammapy:2019}, the standard science tool for CTA. Background estimation is done with the reflected multiple off technique \cite{multipleOFF_berge2007}. The excess and the background are calculated per reconstructed energy bin.

\begin{figure}
     \centering
     \begin{subfigure}[b]{0.48\textwidth}
         \centering
         \includegraphics[width=\textwidth]{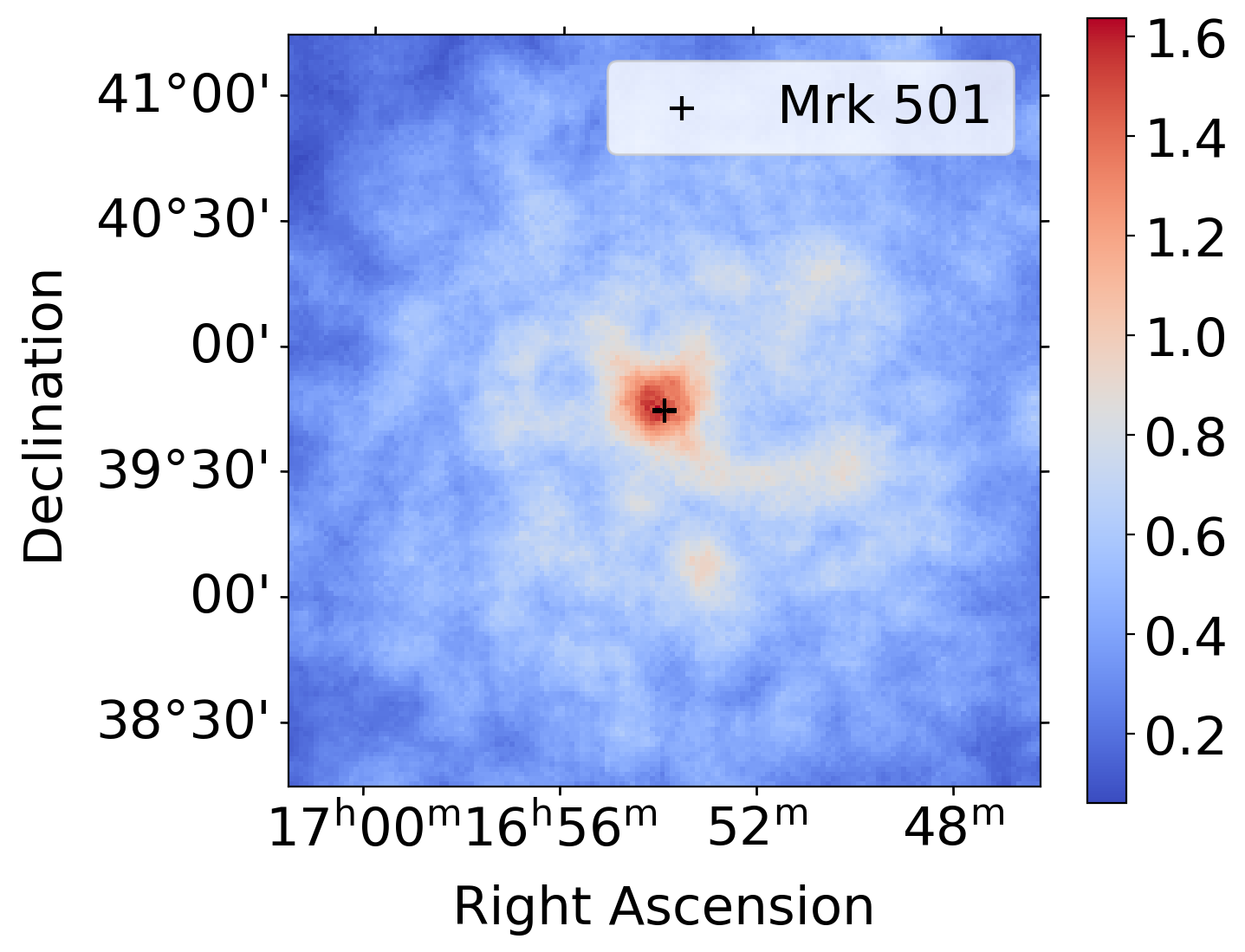}    
         \caption{Hillas+RF}
     \end{subfigure}
     \hfill
     \begin{subfigure}[b]{0.48\textwidth}
         \centering
         \includegraphics[width=\textwidth]{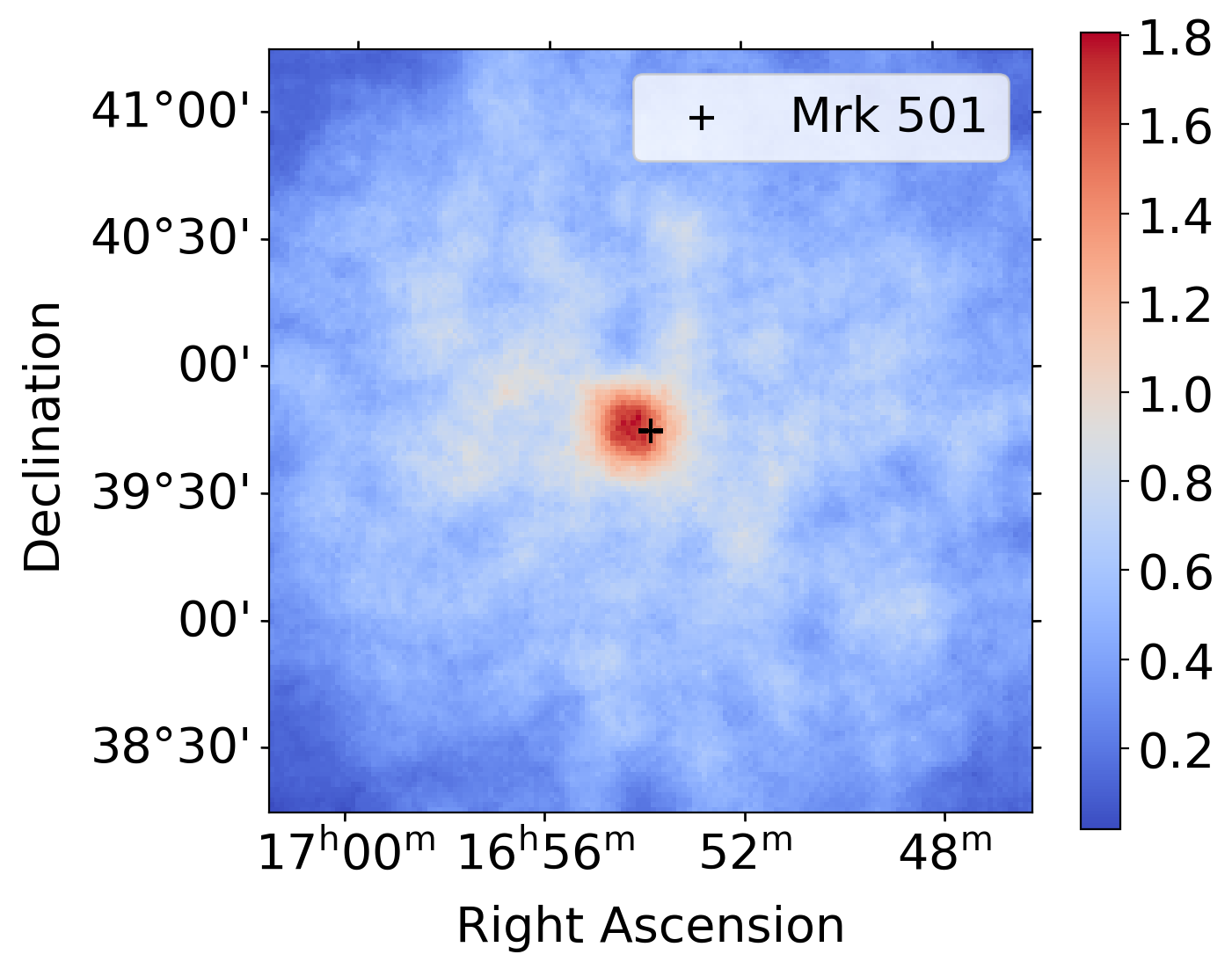}
         \caption{\gphysnet}
     \end{subfigure}
        \caption{Smoothed (kernel=0.12\textdegree) count map of gamma-like events around Markarian 501.\label{fig:CountMap}}
\end{figure}

\begin{figure}[h]
\begin{minipage}[c]{0.49\linewidth}
    \centering
    \includegraphics[width=\linewidth]{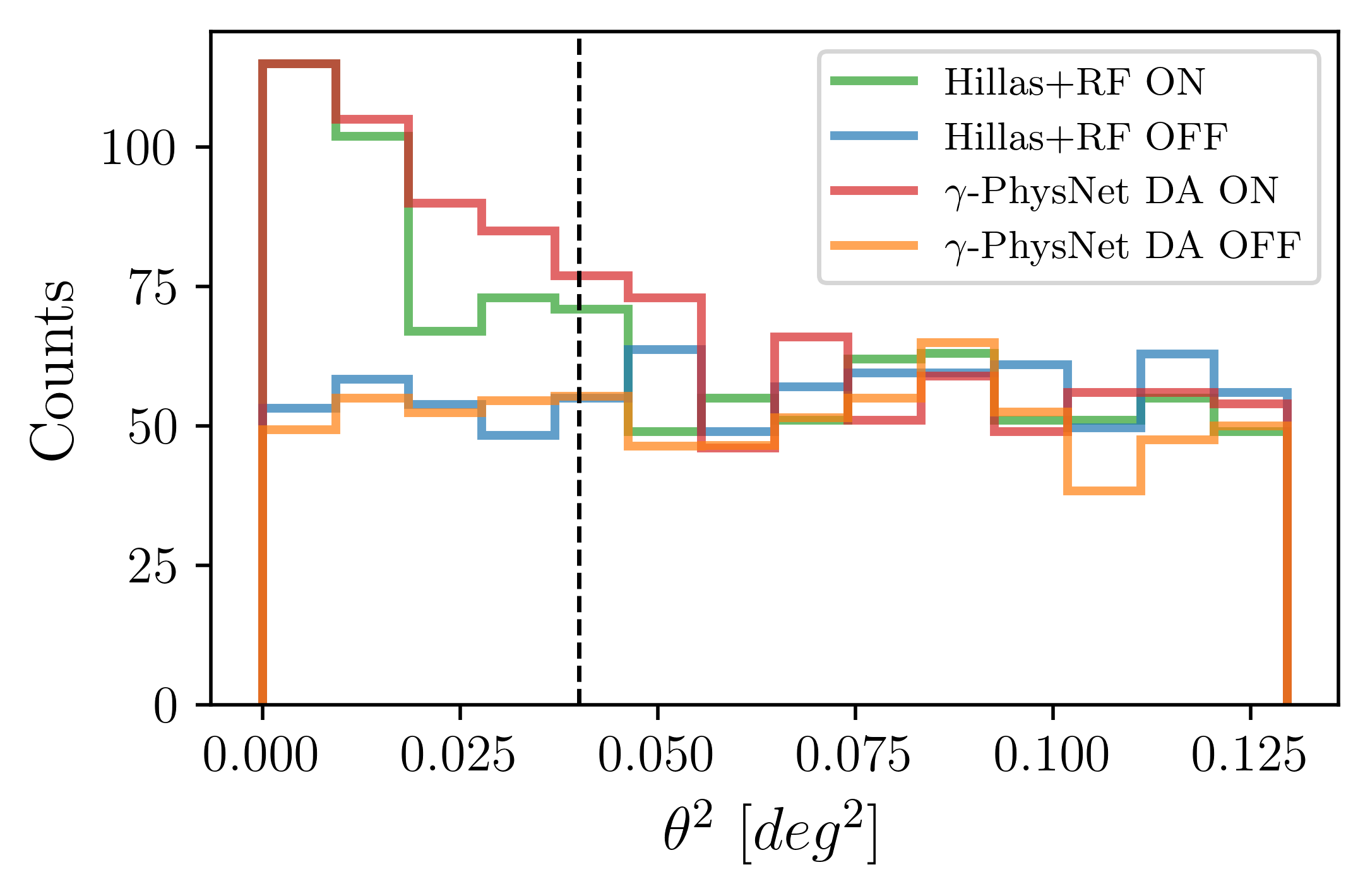}
    \caption{Event count as a function of the squared angular separation to Mrk 501. The vertical dashed line represent the $\theta^2$ cut applied on data for the analysis.}
    \label{fig:theta2_mrk501}
\end{minipage} \hfill
\begin{minipage}[c]{0.49\linewidth}
    \centering
    \includegraphics[width=\linewidth]{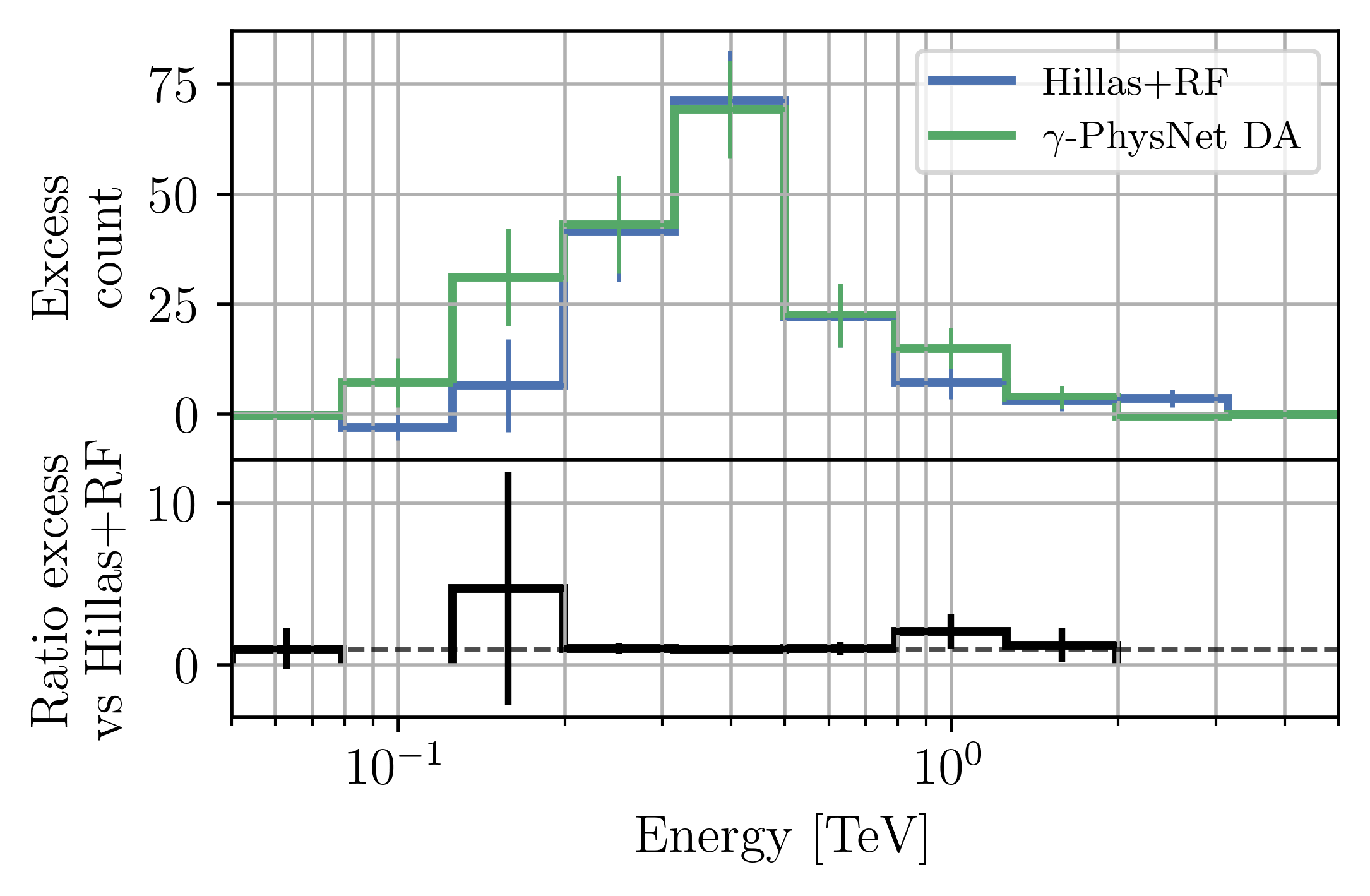}
    \caption{Excess and relative excess between \gphysnet~and Hillas+RF as a function of the events reconstructed energy. Bins are missing in the ratio in case of no positive excess for Hillas+RF.}
    \label{fig:excess_energy_mrk501}
\end{minipage}
\end{figure}

First a counts map of gamma-like events is generated for both chains (Fig. \ref{fig:CountMap}). For \gphysnet~and Hillas+RF we can notice a hot spot at the source position, later confirmed by proper background estimation and significance calculation. Figure \ref{fig:theta2_mrk501} shows the events counts as a function of the separation angle to the source for both chains in OFF and ON regions. On these figures one can observe a bias in the direction reconstruction in the case of \gphysnet. The reason of this bias on observations needs further investigation but preliminary studies show an impact of the discrepancy in pointing direction between simulations and observations. An angular cut of $\theta=0.2\deg$ is applied on data to select the source region in agreement with the average angular resolution and standard cuts applied in LST-1 data analysis \cite{cta_perf_icrc2021}.

Figure \ref{fig:excess_energy_mrk501} presents a comparative excess count as a function of the reconstructed energy. We can notice the greater excess of gamma-like events detected by \gphysnet~compared to the standard Hillas+RF approach, especially at low energies (around \SI{100}{\giga\electronvolt}). This is in harmony with the results obtained on Monte Carlo simulations presented in figure \ref{fig:irfs}, even though the performance gain is not as large.

Table \ref{tab:excess_mrk501}  provides the integrated counts and significance for both analysis chains. We can notice that the background levels are similar, as imposed by the procedure. We show here that both analysis chains are able to detect Mrk501, with a slight advantage for \gphysnet, showing an integrated excess greater by 30 \%  ($\pm 23\%$) and a higher significance than the reference.

\begin{table}[h]
    \centering
    \begin{tabular}{|c|c|c|c|}
        \hline
         Reconstruction & Excess & Significance & Background counts \\ \hline
         Hillas+RF & 148.7 $\gamma$ & 7.6 $\sigma$ & 238.3 \\
         \gphysnet  & 192.7 $\gamma$ & 9.8 $\sigma$ & 226.3 \\
         \hline
    \end{tabular}
    \caption{Excess and significance results for Markarian 501.}
    \label{tab:excess_mrk501}
\end{table}

\subsection{Detection of the Crab Nebula}

Here we analyze two observation runs, \#2012 (OFF run) and \#2013 (ON run) taken in February 2020 at zenith angles of $21.4$\textdegree\ and $28.9$\textdegree~respectively. Both runs undergo the same data reduction as described in section \ref{sec:mc_perf}.


\begin{wrapfigure}{r}{0.45\textwidth}
  \begin{center}
  \vspace{-2\baselineskip}
    \includegraphics[width=0.45\textwidth]{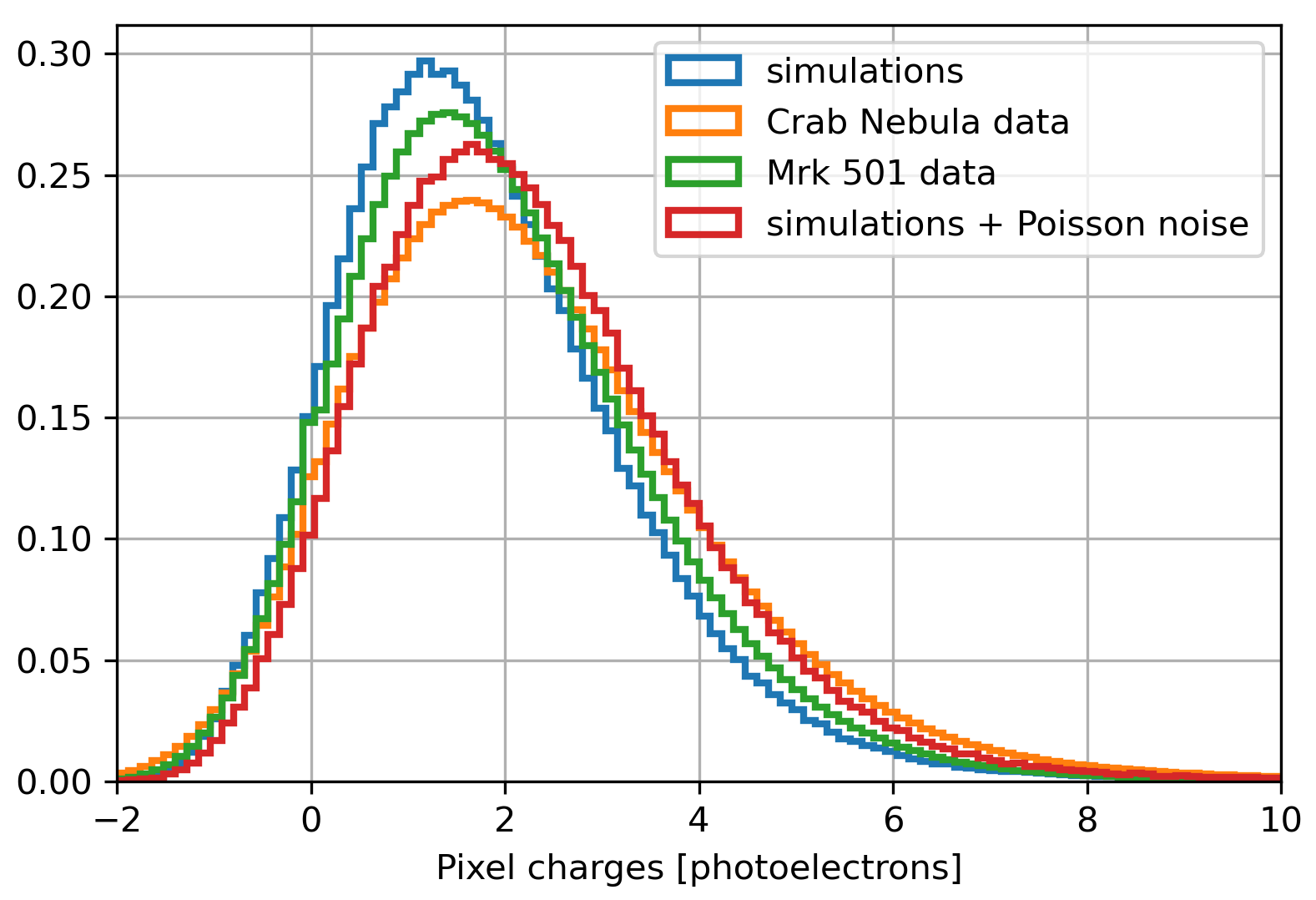}
  \end{center}
  \caption{Distribution of the charges (in photoelectrons) from NSB in the simulations, compared to the one from LST-1 observation data.}
  \label{fig:noise_diff}
  \vspace{-0.2cm}
\end{wrapfigure}

The night sky background (NSB) in the Crab Nebula sky region is much higher than in the case of Mrk 501 and more importantly, much higher than the standard NSB used in simulations. As this can negatively impact the analysis, especially in the case of CNNs which are very sensitive to the noise distribution in the images, we add a poissonian noise to the simulated data. Thanks to Raikov's theorem \cite{raikov1937decomposition}, we determine the parameter of the Poisson distribution as the difference between the simulated average noise pixel charge and the observed one. We can see that the additional noise leads to a better agreement between simulated and observed NSB distributions in figure \ref{fig:noise_diff}. All models are then retrained with these modified simulations.

Here the standard \textit{Hillas+RF} (without added noise) serves as a reference for the rest of the analysis and is used for background matching to all others analysis as described in section \ref{sec:methodo}.

\begin{figure}[h]
\begin{minipage}[c]{0.49\linewidth}
    \centering
    \includegraphics[width=\linewidth]{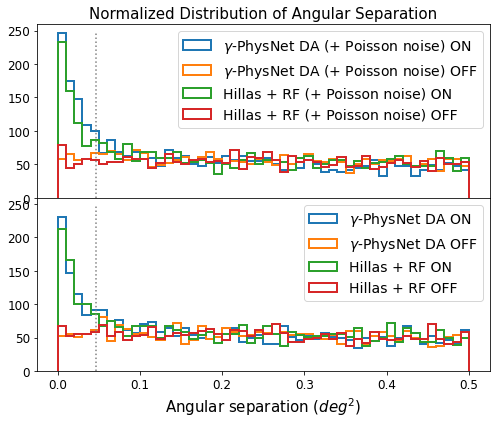}
    \caption{Gamma-ray event distribution in both observations as a function of the squared angular separation between the reconstructed and the true directions.}
    \label{fig:theta2_crab}
\end{minipage} \hfill
\begin{minipage}[c]{0.49\linewidth}
    \centering
    \includegraphics[width=\linewidth]{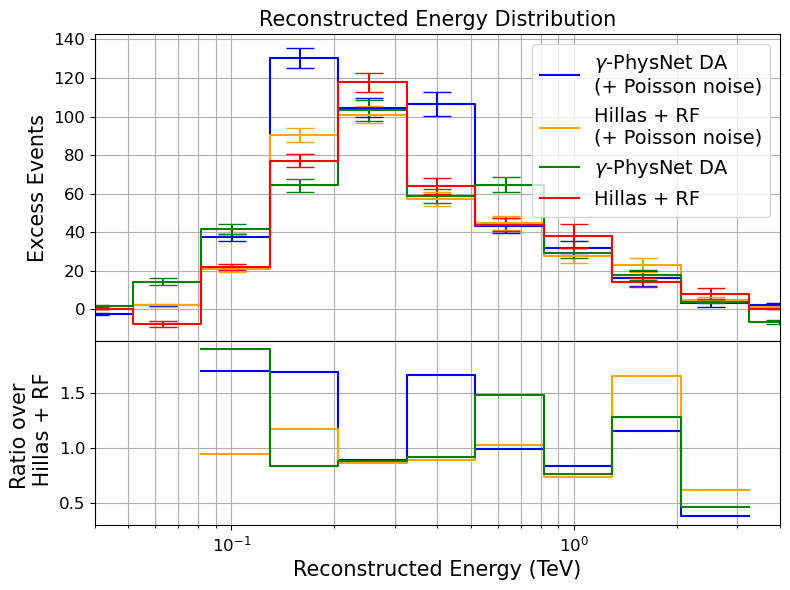}
    \caption{Excess of detected gamma rays per energy bin. The lower part of the plot represents the ratio of gamma-ray excess per energy bin detected by the models over the one detected by the baseline Hillas+RF method.}
    \label{fig:excess_energy_crab}
\end{minipage}
\end{figure}

Figure \ref{fig:theta2_crab} presents the normalized distributions of the angular separation and figure \ref{fig:excess_energy_crab} the excess for all analysis chains. One can see here again the higher excess detected by \gphysnet~compared to Hillas+RF, especially at low energies.

Table \ref{tab:crab_excess} displays the final results on Crab Nebula with the excess count, background count and significance. There are several things worth noticing here. First, one can see the higher significance of \gphysnet~in its first version (12.5 $\sigma$) compared to Hillas+RF (12.0 $\sigma$). Then one can see the important improvement brought by the addition of poissonian noise to the simulated dataset, with a final significance of 14.3 $\sigma$.
Finally, one can note the higher robustness of the Hillas+RF analysis, obtaining very similar results (significance of 11.9 $\sigma$) after the addition of noise. This is most probably due to the tailcut cleaning applied in this chain which is removing the NSB, thus trading loss of information with robustness.

\setlength{\tabcolsep}{10pt}
\begin{table}[bt]
\begin{center}
\begin{tabular}{|c|c|c|c|}
\hline
 & \textbf{Excess}& \textbf{Significance} & \textbf{Background} \\
\hline
Hillas+RF & 379 & $12.0\, \sigma$ & 308 \\
\hline
Hillas+RF & 376 & $11.9\, \sigma$ & 305 \\
\textit{+ Poisson noise} & & &  \\
\hline
$\gamma$-PhysNet DA& 395 & 12.5\, $\sigma$ & 302\\
\hline
$\gamma$-PhysNet DA& \textbf{476} & \textbf{14.3\, $\sigma$} & 317\\
\textit{+ Poisson noise} & & &  \\
\hline
\end{tabular}
\caption{Excess and significance results for Crab Nebula.}
\label{tab:crab_excess}
\end{center}
\end{table}

\section{Conclusion}

After showing better performances on simulated data, the present study confirms the better sensitivity, especially at low energies, of \gphysnet~compared to the classical Hillas+RF reconstruction on observations, with two sources detected with higher significance. The greater gain at low energy can easily be explained by the fact that images are on average noisier at these energies and that convolutional neural networks are more powerful at extracting signal in this case. We also show the importance of closely matching the simulated data with the observed ones taking the NSB as use-case, especially in the case of CNNs which are more sensitive to differences between training and test data. The difference in pointing direction between training data and observation is also expected to play a role and its impact on performances will be studied in a later work.
This work shows that CNNs are a valid approach for LST-1 event reconstruction, even though it is still in commissioning phase. As shown, their performances can still be improved by a closer match between simulations and data, which will be further refined by our increased understanding of the telescope.
\noindent We gratefully acknowledge support from the agencies and organizations listed here: \url{www.cta-observatory.org/consortium\_acknowledgment} and here: \url{https://gammalearn.pages.in2p3.fr/pages/acknowledgements/}.


\setlength{\bibsep}{0.0pt}
\bibliographystyle{IEEEtran}
\bibliography{references}

\clearpage
\section*{Full Authors List: \Coll}

\scriptsize
\noindent

H. Abe$^{1}$,
A. Aguasca$^{2}$,
I. Agudo$^{3}$,
L. A. Antonelli$^{4}$,
C. Aramo$^{5}$,
T.  Armstrong$^{6}$,
M.  Artero$^{7}$,
K. Asano$^{1}$,
H. Ashkar$^{8}$,
P. Aubert$^{9}$
% ...
\\

\noindent
$^{1}$Institute for Cosmic Ray Research, University of Tokyo.
$^{2}$Departament de Física Quàntica i Astrofísica, Institut de Ciències del Cosmos, Universitat de Barcelona, IEEC-UB.
$^{3}$Instituto de Astrofísica de Andalucía-CSIC.
$^{4}$INAF - Osservatorio Astronomico di Roma.
$^{5}$INFN Sezione di Napoli.
$^{6}$Aix Marseille Univ, CNRS/IN2P3, CPPM.
$^{7}$Institut de Fisica d'Altes Energies (IFAE), The Barcelona Institute of Science and Technology.
$^{8}$IRFU, CEA, Université Paris-Saclay.
$^{9}$LAPP, Univ. Grenoble Alpes, Univ. Savoie Mont Blanc, CNRS-IN2P3, Annecy.
% ...


%
%
%

\end{document}